\documentclass[conference,final]{IEEEtran}
\IEEEoverridecommandlockouts

\usepackage[utf8]{inputenc}
\usepackage[T1]{fontenc}
\usepackage[english]{babel}
\usepackage{microtype}
\usepackage{blindtext}

\usepackage{caption}
\usepackage{footnote}
\makesavenoteenv{tabular} 
\usepackage{sepfootnotes}
\sepfootnotecontent{noMulticast}{Lower bound that does not include multicast event routing and assumes physical transmission delay of \SI{4}{\nano\second}. Note this latency does not include the routing chiplet to \gls{bss2} chiplet latency.}
\sepfootnotecontent{asyncErrorcontrol}{The SpiNNaker-1 and DYNAP-SE2 \gls{c2c} links operate asynchronously making them less susceptible to random bit errors.\label{asyncerror}}
\sepfootnotecontent{joschac2c}{Hop latency from the \gls{fpga} that controls a \gls{bss2} chip to the aggregator \gls{fpga}.}
\sepfootnotecontent{spinnlinkb2b}{Board to board latency between two spiNNlink \gls{fpga}s, not the end to end latency from one SpiNNaker core to the next.}
\sepfootnotecontent{extollLatency}{This includes only the EXTOLL derived single hop latency, not the added latency of the event packing scheme.}
\sepfootnotecontent{packetEventSize}{Does not include packet header overhead.\label{noheader}}

\newcommand{\mycitestyle}{ieee}
\IfFileExists{ieee-comp.cbx}{
	\renewcommand{\mycitestyle}{ieee-comp}
}{}
\usepackage[
	style=ieee,
	citestyle=\mycitestyle,
	backend=biber
]{biblatex}
\usepackage{makecell}
\usepackage[pdftex]{graphicx}
\graphicspath{{../../fig/cropped}{../../build/figures}}

\usepackage{pgfplots}
\pgfplotsset{compat=1.16}

\usepackage[acronym]{glossaries}
\newglossaryentry{asic}
{
	name=ASIC,
	description={ASIC}
}

\newglossaryentry{bss2}
{
	name=\mbox{BrainScaleS-2},
	description={BSS-2 system}
}

\newglossaryentry{bss1}
{
	name=\mbox{BrainScaleS-1},
	description={BSS-1 system}
}

\newglossaryentry{fpga}
{
	name=FPGA,
	description={FPGA}
}

\newglossaryentry{in}
{
	name=IN,
	first={Interconnection Network},
	description={Interconnection Network}
}

\newglossaryentry{pcb}
{
	name=PCB,
	description={Printed Circuit Board}
}

\newglossaryentry{soc}
{
	name=SoC,
	description={System on Chip}
}

\newglossaryentry{adc}
{
	name=ADC,
	description={Analog Digital Converter}
}

\newglossaryentry{c2c}
{
	name=C2C,
	description={Chip to Chip}
}

\newglossaryentry{ddr}
{
	name=DDR,
	description={Double Data Rate}
}

\newglossaryentry{phit}
{
	name=phit,
    first={physical transfer unit},
	description={physical transfer unit}
}

\newglossaryentry{flit}
{
	name=flit,
	description={flow control unit}
}

\newglossaryentry{ppa}
{
	name=PPA,
	description={Power, Performance and Area}
}

\newglossaryentry{fec}
{
	name=FEC,
	description={Forward Error Correction}
}

\newglossaryentry{crc}
{
	name=CRC,
	description={Cycling Redundancy Checksum}
}

\newglossaryentry{hd}
{
	name=HD,
	description={Hamming Distance}
}

\newglossaryentry{arq}
{
	name=ARQ,
	description={Automatic Repeat-reQuest}
}

\newglossaryentry{aer}
{
	name=AER,
	description={Address Event Representation}
}

\newglossaryentry{ocp}
{
	name=OCP,
	description={Open Core Protocol}
}

\newglossaryentry{axi}
{
	name=AXI,
	description={Advanced eXtensible Interface}
}

\newacronym{lif}{LIF}{leaky-integrate-and-fire}
\newacronym{snn}{SNN}{spiking neural network}
\newacronym{ppu}{PPU}{plasticity processing unit}
 
\usepackage{booktabs}
\usepackage{nicefrac}
\usepackage[obeyFinal]{todonotes}
\usepackage{csquotes}

\usepackage{siunitx}
\usepackage{tikz}
\usepackage{circuitikz}
\usetikzlibrary{decorations.pathreplacing,decorations.text}
\usetikzlibrary{calc}
\usetikzlibrary{fit}

\tikzstyle{panel} = [
inner sep=0pt,
]
\tikzstyle{panellabel} = [
anchor=north west,
inner sep=0pt,
font={\bfseries\sffamily},
]

\usepackage{listings}
\usepackage{algorithm}
\usepackage{algpseudocodex}
\usepackage{enumitem}

\PassOptionsToPackage{hyphens}{url}
\usepackage[capitalise,nameinlink]{cleveref}

\newcommand{\cpp}{C++}

\renewbibmacro{finentry}{\iffieldequalstr{entrykey}{rijpkema2003tradeoffs}{\finentry\newpage}
	{\finentry}}

\newcommand\submittedtext{\footnotesize
	\textcopyright\,2026 IEEE\@.
	Personal use of this material is permitted.
	Permission from IEEE must be obtained for all other uses, in any current or future media, including reprinting/republishing this material for advertising or promotional purposes, creating new collective works, for resale or redistribution to servers or lists, or reuse of any copyrighted component of this work in other works.
}
\newcommand\submittednotice{\begin{tikzpicture}[remember picture,overlay]
	\node[anchor=south,yshift=30pt] at (current page.south) {\fbox{\parbox{\dimexpr0.82\textwidth-\fboxsep-\fboxrule\relax}{\submittedtext}}};
	\end{tikzpicture}}

\begin{document}

\title{
  A Unified Interconnection Network for Chiplet-Based Scaling of the BrainScaleS Neuromorphic System
}

\author{\IEEEauthorblockN{Robin Heinemann\IEEEauthorrefmark{3}
        and
		Johannes Schemmel\IEEEauthorrefmark{3}
	}
	\vspace{2mm}

	\IEEEauthorblockA{\IEEEauthorrefmark{3}\,Institute of Computer Engineering, Heidelberg University, Heidelberg, Germany
	}\vspace{-2mm}

	\thanks{
		The presented work used systems, which received funding from 
		the European Union's Horizon 2020 Framework Programme for Research and Innovation under the Specific Grant Agreements Nos. 720270, 785907 and 945539 (Human Brain Project, HBP) and Horizon Europe grant agreement No. 101147319 (EBRAINS 2.0)

		{\setlength{\parskip}{2pt}
			\noindent Correspondence: robin.heinemann@ziti.uni-heidelberg.de
		}
	}}

\maketitle

\submittednotice

\begin{abstract}
	The BrainScaleS-2 (BSS-2) neuromorphic architecture combines analog emulation of spiking neural network (SNN) primitives with tightly coupled ADCs and digital processing units.
These analog SNN primitives are fixed hardware resources that cannot be multiplexed, limiting the emulated network size to the number of physical hardware copies.
To overcome the challenges of scaling analog designs, chiplet-based designs offer a promising approach with cost and flexibility advantages over monolithic scaling.
Implementing a chiplet-based BSS-2 architecture requires an interconnection network that handles two distinct classes of traffic: error-tolerant traffic such as spikes and error-intolerant traffic like configuration data or data exchanged by the processing units.
This work presents the design of a routing chiplet for the BSS-2 architecture that enables interconnection of multiple BSS-2 units in a 2D mesh topology.
Both traffic classes are multiplexed over a single wide parallel die-to-die link. Exploiting the fault tolerance of SNNs, spikes are transmitted unsecured and synchronously, with the arrival time on the receiving side directly determining the pre-synaptic time of the spike.
Conversely, error-intolerant data transmission is secured by a point-to-point Automatic Repeat Request protocol and uses credit-based flow control.
These design choices are validated in simulation, where the proposed design can sustain 95\% link bandwidth utilization for the use case of surrogate gradient training across a wide range of spike-to-secured traffic ratios under a \num{1e-10} bit error rate with little impact on spike timing jitter.
 \end{abstract}

\section{Introduction}\label{sec:introduction}

\begin{figure}[ht!]
	\vspace{-2mm}
	\centerline{
		\includegraphics{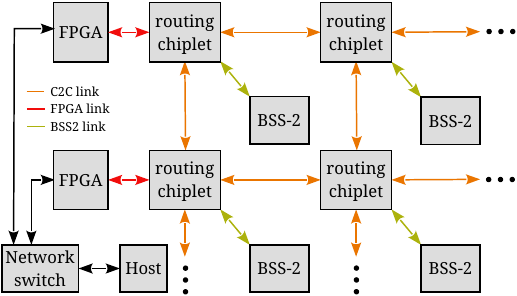}
	}
	\caption{
	  Schematic overview of a chiplet-based scaling of the \gls{bss2} neuromorphic system. A 2D Mesh of routing chiplets implements an Interconnection Network (\gls{in}), and each routing chiplet can be attached to a mixed signal \gls{bss2} \gls{asic}. \gls{fpga}s connecting to some of the routing chiplets allow control of the system from a conventional host computer.
	}
	\label{fig:mesh_schematic}
\end{figure}

\begin{figure*}[ht]
	\centerline{
		\includegraphics{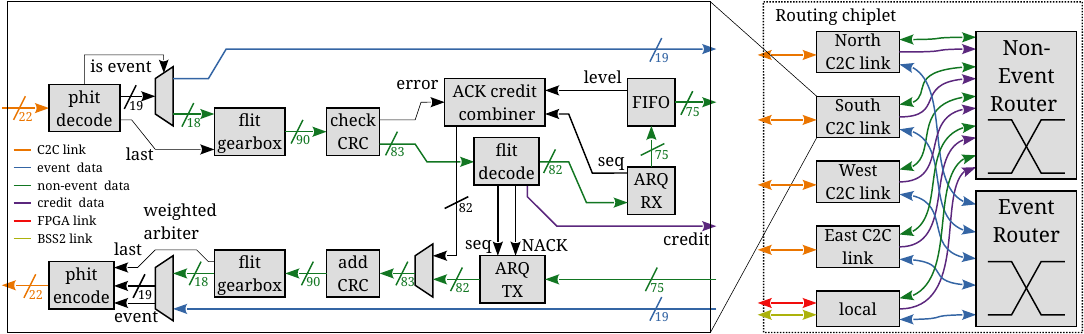}
	}
	\caption{
      Schematic overview of the routing chiplet architecture. Incoming \gls{phit}s get decoded into event or non-event message payload, with multiple non-event \gls{phit}s getting accumulated using a framing bit into a larger \gls{flit}, which carries a \gls{crc} used to detect any payload errors. An \gls{arq} scheme ensures retransmission of any \gls{flit}s received with an error. Correctly received \gls{flit}s get cumulatively acknowledged using a special acknowledgement \gls{flit} that gets transmitted using the outgoing data path. The same \gls{flit} type also carries credit information used for flow control. Outgoing \gls{flit}s travel the inverse path, they get buffered by the outgoing \gls{arq} logic until their successful reception is acknowledged and get split up into \gls{phit}s for transmission over the physical link. Event messages get transmitted over the same physical link with a weighted arbitration scheme that by default prioritizes event messages. Non-event messages get routed from a given input port to their target output port by a wormhole router using dimension-ordered X-Y routing. The event router design is not part of this work. For evaluation purposes a simple broadcast event router was included.
      \vspace{-3mm}
	}
	\label{fig:detail_view}
\end{figure*}

Chiplet-based \gls{soc}s offer a promising alternative to large monolithic \gls{soc}s. They can offer a cost advantage by optimizing yield and power or performance binning~\parencite{kannan2015disintegration}.
Furthermore, by splitting up a large monolithic die into multiple smaller ones, they allow for more modularity and easier design space exploration, by allowing parts of an \gls{soc} to be exchanged at a finer granularity. The ability to choose different process nodes for different parts of the system conveys further advantages, especially for analog or mixed-signal designs that do not benefit from the digital compute power of the newest process nodes.
One example is the BrainScaleS neuromorphic system. 
Its current in-silico realization, the \gls{bss2} \gls{soc}, targets a \SI{65}{\nano\meter} process. It consists of an analog core that performs continuous-time emulation of up to \num{512} neurons and \num{131072} synapses combined with on-chip, parallel ADCs and a tightly coupled embedded processor~\parencite{pehle2022brainscales2}. \gls{bss2} operates at an acceleration factor of \(\num{1000}\times\) compared to biological real-time and the embedded processor allows for flexible emulation of complex plasticity rules~\parencite{atoui2025multi}. The analog core is area-constrained and does not benefit from newer process nodes in the same way that the embedded processor can, making chiplet-based scaling of the BrainScaleS system a compelling design objective.
Data exchanged between the neurons takes the form of spikes that get emitted by a neuron once it reaches its activation potential. A spike carries no payload; instead information is conveyed in the timing of spikes. Drawing inspiration from biology, where synaptic transmission is unreliable~\parencite{hessler1993probability} a neuromorphic system is expected to operate correctly even in the presence of a small loss of spikes.
The analog emulation of the neurons and synapses operates in continuous-time and cannot be stopped, which means that a scaled-up system has to operate in a globally asynchronous fashion. Furthermore, congestion will alter spike timing, so it is expected that congested spikes can be dropped.
This reasoning, however does not hold for data like configuration data or messages exchanged by the embedded processors, where reliability and correctness are essential for the system to be usable.

A chiplet-based scaled-up \gls{bss2} system therefore needs a unified \glsfirst{in} (IN) that can carry messages of two different classes 
\begin{enumerate}
\item event messages, that are forwarded with minimal and deterministic latency to preserve their relative timing, but that can tolerate a small loss of messages due to congestion or transmission errors
\item non-event messages, for which reliability and correctness are guaranteed
\end{enumerate}

This work presents the design of such a unified \glsfirst{in} (IN). In the following sections, the architecture of the proposed \gls{in} is described and the design choices are validated in simulation for some basic operation modes.
 \section{Interconnection Network description}\label{sec:methods}
\cref{fig:mesh_schematic} shows the high-level system overview. Dedicated routing chiplets are connected in a 2D mesh and implement the \gls{in}. Each routing chiplet can be connected to a \gls{bss2} \gls{soc} or an \gls{fpga}, which allows control from a conventional computer. Previous work has shown that the most energy-efficient error control scheme depends on the physical link's parameters, including the link latency, the bit error rate and the bandwidth of the links\parencite{bertozzi2005error,murali2005error,ejlali2010error}. The design choices for this \gls{in} therefore must be considered in the context of the prospective physical link used. This work considers an interconnection network built using the \gls{c2c} links described in \parencite{ilmberger2024flexible}, which are tested to \(\leq \num{1e-10}\) bit error rate at a data rate of \SI{2}{\giga\bit\per\second}. Bonding and area constraints limit the number of links carrying data per direction to about \num{11}, for a total bandwidth of \SI{22}{\giga\bit\per\second} per direction and port. 

In the following, the architecture for the \gls{in} is described in detail and \cref{fig:detail_view} gives a schematic overview of it.. 
For each port, the \gls{c2c} links are organized as a parallel bus. They operate at \gls{ddr} transmitting a \SI{22}{\bit} \gls{phit} every \SI{1}{\nano\second} clock cycle.
Schemes that partition the available link bandwidth statically to specific message classes suffer under uneven traffic loads~\parencite{rijpkema2003tradeoffs}. Therefore, a dynamic partitioning scheme is used here, where event messages are prioritized and non-event messages use the rest of the available bandwidth. To achieve this, event and non-event traffic flows are split on the \gls{phit} level, with each \gls{phit} including a bit indicating its type.
For outgoing \gls{phit}s a weighted priority arbiter arbitrates between \gls{phit}s from event traffic and non-event traffic. The arbiter guarantees a configurable minimum throughput of non-event-type messages and prioritizes event-type \gls{phit}s beyond that. Multiple non-event \gls{phit}s get combined into fixed-sized \gls{flit}s. For synchronization of the gearboxes on the sending and receiving side that split up / accumulate the \gls{phit}s into \gls{flit}s each \gls{phit} includes a framing bit. While in principle it would be possible to synchronize these gearboxes without any additional overhead by for example relying on the \gls{crc} included in the non-event messages,
including a dedicated framing bit in each \gls{phit} allows for fast recovery from a \gls{phit} transmitted with errors that would cause the gearboxes to get out of sync. Here the meaning of the framing bit is chosen to indicate that a \gls{phit} is the last \gls{phit} of a non-event message.
Misinterpretation of a \gls{phit} type or the framing bit for non-event traffic is expensive to protect against with a \gls{crc}, as a misinterpretation is equivalent to a random bitflip for all bits of a \gls{phit}.
The type and framing bit are therefore protected against single bit flips with a Hamming-(7,4) \gls{fec} code. Overhead of this \gls{fec} code is minimized by grouping \gls{phit}s into pairs with a shared header, bringing the overhead to \SI{3}{\bit} for event-type \gls{phit}s and \SI{4}{\bit} for non-event-type \gls{phit}s as shown in \cref{tab:phitCoding}. 

\begin{table}[t]
  \begin{minipage}{\columnwidth}
\begin{center}
    \caption{Header encoding used for a pair of \gls{phit}s\label{tab:phitCoding}}
\begin{tabular}{lll}
  \toprule
payload type 1 & payload type 2 & encoding \\
  \midrule
event & event & ${111000}_2$ \\

event & none & ${1001100}_2$ \\
event & non-event & ${0111100}_2$ \\
event & non-event last & ${0101010}_2$ \\

non-event & event & ${1011010}_2$ \\
non-event & non-event & ${1101100}_2$ \\
non-event & non-event last & ${0010110}_2$ \\

non-event last & event & ${1000011}_2$ \\
non-event last & non-event & ${0001111}_2$ \\

none & none & ${0000000}_2$ \\
\bottomrule
\end{tabular}
\end{center}
  \vspace{2.8mm}
  \end{minipage}
Optimized encoding for \gls{phit} header. The constructed prefix code only requires \SI{6}{\bit} to encode a pair that contains two events, while all other pairs require \SI{7}{\bit}. Each code differs by at least three bits from every other code, so a single bit flip can be corrected.
\end{table}

Using \gls{aer}\parencite{mahowald1992phd}, the resulting \SI{19}{\bit} for the event payload allows a unique label for up to \(2^{19}\) = \num{524288} neurons or at least \num{1024} \gls{bss2} chips. For real systems it is expected that each chip only sees event messages from a subset of the neurons in the whole system, allowing for non-unique labels, increasing the maximum possible system size.

Error detection and control for the payload of non-event-type messages is performed on a per-\gls{flit} level, with each \gls{flit} including a \gls{crc}. For the proposed system each \gls{flit} is composed of \SI{90}{\bit} allowing protection for a Hamming distance of \num{3} using only \SI{7}{\bit} for the \gls{crc}.

The error control scheme follows a link-level Automatic repeat request (\gls{arq}) scheme optimized for minimal latency impact of recovery from an error. Each \gls{flit} contains a sequence number and successful reception of a \gls{flit} is communicated to the sender with a special acknowledgement \gls{flit}. 
Errors detected by the \gls{crc} are remedied by retransmission of the \gls{flit} by the sender, which is either triggered by reception of a special non-acknowledgement \gls{flit} that the receiver sends out when it detects an error or by a timeout if there are any outstanding unacknowledged \gls{flit}s.

Flits use credit-based flow control to transmit backpressure across chip boundaries. The credit count is hereby transmitted as part of the special acknowledgement \gls{flit}s.

Routing of the \gls{flit}s is performed in a wormhole fashion, with one or more \gls{flit}s combining into a larger packet. An initial \gls{flit} carries the routing information used to determine the target output port for the packet. The \gls{in} and router impose no artificial limit on the length of a packet, minimizing the overhead induced by the routing information.

Tunneling of protocols with ordering guarantees like \gls{axi} or the \gls{ocp}~\parencite{ocp30} employed on the \gls{bss2} \gls{asic} is simplified by choosing a deterministic routing function, in this case dimension-ordered X-Y routing, which guarantees that the interconnection network preserves the order of packets between every source target pair. No reordering buffer on the receiving side is required. Note, however, that if a source allows for multiple outstanding reads to different targets it still needs a reordering buffer for read response data, as there is no ordering guarantee for the read responses sent by different targets to the same source.

The event router is not considered as part of this work. For evaluation purposes a simple broadcast event router was included.

 \section{Simulation-based evaluation}\label{sec:results}
To validate the design choices of the proposed \gls{in} its behavior is investigated in simulation.
The described components, phit encoders / decoders, \gls{arq} implementation and non-event router were implemented in synthesizable SystemVerilog or Amaranth HDL~\parencite{amaranth}. Their correctness was verified using extensive formal verification. To investigate the behavior under errors of the presented \gls{in} this implementation was compiled into a cycle-accurate \cpp{} model using CXXRTL~\parencite{cxxrtl} and combined with high-level models of the physical links, event and non-event traffic implemented in \cpp{}.

\begin{figure}
  \vspace{2pt}
	\begin{tikzpicture}
		\node[panel, anchor=north west] (a) at (0, 0) {
			\input{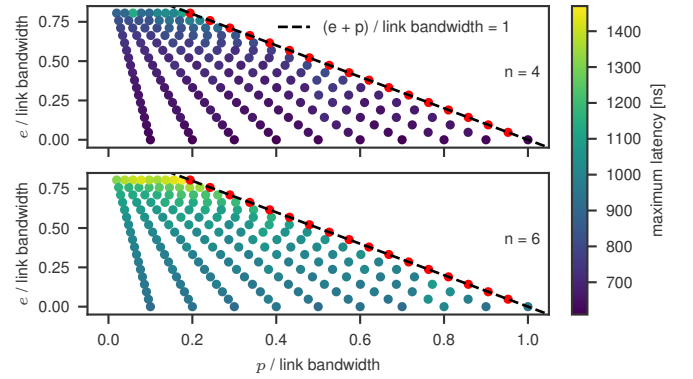}
		};
	\end{tikzpicture}
	\vspace{-6mm}

	\caption{Maximum latency for non-event messages to reach their target for different event message rates $e$ and non-event message rates $p$ for different number of routing chiplets $n$ in a setup, as shown in \cref{fig:mesh_schematic}. Events are generated with a Bernoulli process while non-event messages get injected at constant rate. Red marks cases where the imposed message rates was not accepted. The links were simulated to be error free.
		\vspace{-3mm}
	}
	\label{fig:lat}
\end{figure}
 
For the physical links a fixed-latency lossy channel with a fixed bit error rate is simulated. Even though the non-event messages trade throughput and latency for reliability and correctness, due to the continuous time nature of the larger system, the bandwidth and latency achievable in this \gls{in} for non-event messages are still of interest. First, the maximum latency for non-event messages to reach their target in a setup, as shown in \cref{fig:mesh_schematic}, is evaluated.
The traffic pattern chosen for this evaluation is designed to mimic hardware-in-the-loop surrogate gradient training.
Surrogate gradients are a promising technique for training \glspl{snn}. They can be applied to \gls{bss2} by operating in a hardware-in-the-loop fashion, executing the forward pass on the analog computing substrate and using the on chip \gls{adc}s to capture membrane traces, which are used as input to the backward pass executed on a host computer~\parencite{cramer2022surrogate}. 
This gives a uniform message rate for the non-event messages (the \gls{adc} samples). Conversely the event messages are generated from a Bernoulli process for each \gls{c2c} link. \cref{fig:lat} shows the maximum latency for $n = 4$ or $6$ routing chiplets in row, with the bandwidth to the \gls{fpga} assumed to be unlimited. The bit error rate of the physical links is simulated at $0$, and the physical transmission delay is simulated at a large \SI{135}{\nano\second} to test the limits of the \gls{in}. The input buffer size and the ARQ sender size is \num{64} \gls{flit}s each in this case. Parameters where the imposed non-event message rate could not be sustained are shown in red.
For low event message rates, the latency is close to the physical transmission delay, with a minimum of \SI{610}{\nano\second} in the case of n = 4. With increasing event message rate the latency increases due to congestion and the prioritization of event messages.

\begin{figure}
  \vspace{2pt}
	\begin{tikzpicture}
		\node[panel, anchor=north west] (a) at (0, 0) {
          \includegraphics[trim=0pt -0mm 0pt -1mm ]{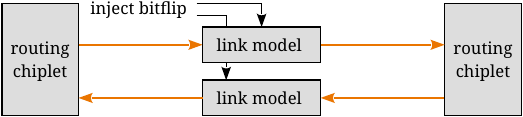}
		};

		\node[panel, anchor=north west] (b) at (0, -2.3) {
			\input{nrecover.pgf}

		};

		\node[panellabel, anchor=west] at (a.north west) {A\vphantom{bp}};
		\node[panellabel, anchor=west] at (b.north west) {B\vphantom{bp}};
	\end{tikzpicture}
	\vspace{-6mm}

	\caption{(A) Schematic overview of the simulated system. The routing chiplets are simulated cycle accurate, while the links simulated as simplified fixed latency lossy links.
		(B) Mean time between failure for different event message rates $e$ and non-event message rates $p$. Events are generated with a Bernoulli process; non-event messages get injected at constant rate. A failure indicates that the given non-event message rate cannot be accepted and backpressure is generated. Red marks cases where a single bitflip causes backpressure to be generated. Bit error rate is simulated to be \SI{1e-10}{\per\bit}.
		\vspace{-3mm}
	}
	\label{fig:mtbf}
\end{figure}

The second aspect investigated is the sustained bandwidth for non-event messages for a link error rate $\neq 0$. Because transmission errors inevitably trigger retransmission, absolute bandwidth guarantees are impossible. Instead here the mean time between failure (MTBF), where a given bandwidth cannot be sustained is measured. Realistic bit error rates of \(\mathcal{O}(\SI{1e-10}{\per\bit})\) are too small to be extracted from simulation directly. Instead an upper bound for the MTBF is calculated from simulation, by measuring the longest time \(t_{\text{recover}}\) required for the steady state buffer levels to be reached again after a transmission error occurs over \num{200} randomly injected bitflips. An upper bound for the MTBF under the assumption of independent errors is then determined from the MTBF for a second transmission error to occur during this time:
\[\text{MTBF} = \frac{1}{t_{\text{recover}} B^2 \rho^2}\]
Where \(B\) is the link bandwidth and \(\rho\) the bit error rate. For this traffic pattern, the routing does not influence the error behavior, therefore only a single bidirectional \gls{c2c} link is simulated. \cref{fig:mtbf} shows the obtained MTBF in days for different average event and non-event message rates for an ARQ sender and input buffer size of again \num{64} \gls{flit}s each as well as a physical link delay of \SI{85}{\nano\second}. This physical link delay was chosen to be the maximum link delay where a single error can be tolerated without immediately filling up the \gls{arq} buffer at \SI{95}{\percent} link utilization. The lower limit of MTBF that can be tolerated depends on the specific experiment performed with the hardware. Usually, failing to sustain a given bandwidth is not expected fatal. For example for ADC samples it is expected that some amount of lost samples can be tolerated. Note also, that this gives the MTBF for a single link, for a complete system it has to be divided by the number of links. For example a \num{16} by \num{16} mesh contains \num{960} chip to chip links, yielding an MTBF of \(\mathcal{O}{\text{Minutes}}\) in the worst case.

\begin{figure}
  \vspace{2pt}
	\begin{tikzpicture}
		\node[panel, anchor=north west] (a) at (0, 0) {
          \includegraphics[trim=0pt -0mm 0pt +2mm ]{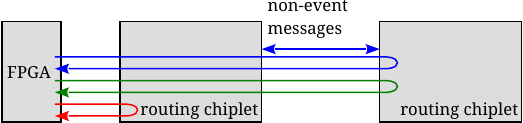}
		};

		\node[panel, anchor=north west] (b) at (0, -2.3) {
			\input{jitter.pgf}
		};

		\node[panellabel, anchor=west, inner sep=0pt] at (a.north west) {A\vphantom{bp}};
		\node[panellabel, anchor=west] at (b.north west) {B\vphantom{bp}};
	\end{tikzpicture}
	\vspace{-6mm}

	\caption{(A) Sketch of the three cases considered for the round trip event latency simulation. 
		(B) Measurement of round-trip latency for \num{10000} event messages in simulation. Red gives the baseline without traversing any \gls{c2c} links, while blue and green traverse two \gls{c2c} links, with non-event messages being transported at maximum rate over the \gls{c2c} link for the blue case. The granularity of the histogram is given by the \SI{125}{\mega\hertz} \gls{fpga} clock used to timestamp the events. 
		\vspace{-3mm}
	}
	\label{fig:jitter}

\end{figure}

Finally the latency and the jitter in the latency incurred by the \gls{in} for event messages are evaluated. For this evaluation end-to-end co-simulation with the current \gls{bss2} hardware revision and \gls{fpga} design was used. Again the physical links were modeled as fixed-latency links. In this case a transmission delay of \SI{4}{\nano\second} was evaluated. No errors were injected, as the event messages do not have error control. 
The interface between \gls{fpga} and \gls{bss2} is reused to connect the routing chiplet to the \gls{fpga}. \cref{fig:jitter} shows the loopback event message latencies for multiple event message rates and traffic scenarios. For this simulation a uniform event message rate was used. In red the baseline where only the \gls{fpga} to routing chiplet link is traversed is shown, while for blue and green two \gls{c2c} links are traversed by the event messages. For blue additional non-event messages are injected onto the link at maximum permissible rate. The jitter for the event messages is dominated by the jitter already introduced by the \gls{fpga} to routing chiplet interface which was reused from the \gls{bss2} \gls{asic}. A thorough investigation of it can be found in \parencite{rettig2019bachelorthesis}. In this simulation a round-trip latency of \SI{32}{\nano\second} between two routing chiplets is measured. 

For an ARQ sender and input buffer size of \num{64}-\glspl{flit} the post-synthesis area estimate of the proposed non-event router, ARQ and link-level protocol in \SI{65}{\nano\meter} technology is \(\approx\)~\SI{0.17}{\milli\meter\squared} with \(\approx\)~\SI{0.13}{\milli\meter\squared} of that being the input and ARQ buffers. Note that this does not include the size of the I/O cells.
 \begin{table*}[ht!]
	\centering
	\caption{Comparison of this work with the \gls{in} of other neuromorphic systems that perform realtime event message transmission. Bandwidth compared is the physical link bandwidth, not the effective payload bandwidth. For systems where latency depends on congestion the uncongested latency is compared.}
	\label{tab:otherSystems}
\begin{minipage}{\textwidth}
\centering
	\begin{tabular}{lccccccc}
		\toprule
		 & \makecell{this work\\\textit{simulated}} & \makecell{\gls{bss2}\\\gls{fpga}~\parencite{ilmberger2025multichip}} & \makecell{\gls{bss2}\\EXTOLL~\parencite{thommes2022demonstrating}} & \makecell{SpiNNaker-1\\\gls{c2c}~\parencite{furber2014spinnaker}} & spiNNlink~\parencite{plana2020spinnlink} & DYNAP-SE2~\parencite{richter2024dynapse2} \\
		\midrule
        \vspace{1mm}\makecell{event\\operation} & realtime & realtime & \makecell{grouped and\\ timestamped} & realtime & realtime & realtime \\
        \vspace{1mm}\makecell{event\\flow control} & none & none & \makecell{credit\\counting} & \makecell{async\\handshake} & \makecell{global credit\\and per channel} & \makecell{async\\handshake} \\
        \vspace{1mm}\makecell{event\\error control} & none & none\textsuperscript{\ref{asyncerror}} & \gls{arq} & none & \gls{arq} & none\textsuperscript{\ref{asyncerror}}  \\
        \vspace{1mm}\makecell{non-event\\flow control} & \makecell{credit\\counting} & \makecell{credit\\counting} & \makecell{not\\implemented} & \makecell{async\\handshake} & \makecell{global credit\\and per channel} & \makecell{async\\handshake} \\
        \vspace{1mm}\makecell{non-event\\error control} & \gls{arq} & \makecell{not\\implemented} & \gls{arq} & none\sepfootnote{asyncErrorcontrol} & \gls{arq} & none\textsuperscript{\ref{asyncerror}} \\
\vspace{1mm}\makecell{transmitted bits\\per event} & 22 & 20 & 29\sepfootnote{packetEventSize} & 40 or 72 & 32 or 64\textsuperscript{\ref{noheader}} & 24 \\
        \vspace{1mm}\makecell{minimum single\\hop latency} & $\geq$\SI{16}{\nano\second}\sepfootnote{noMulticast} & \SI{0.17}{\micro\second}\sepfootnote{joschac2c} & \SI{75}{\nano\second}\sepfootnote{extollLatency} & \SI{0.44}{\micro\second} & \SI{665}{\nano\second}\sepfootnote{spinnlinkb2b} & not reported\\
        \vspace{1mm}\makecell{\gls{c2c} link\\bandwidth} & \SI{22}{\giga\bit\per\second} & \SI{5}{\giga\bit\per\second} & \SI{16}{\giga\bit\per\second} & average \SI{275}{\mega\bit\per\second} & \SI{3}{\giga\bit\per\second} & not reported  \\
\bottomrule
	\end{tabular}
    \vspace{-2mm}
\end{minipage}
    \vspace{-4mm}
\end{table*}

\section{Discussion and Outlook}\label{sec:discussion}
This work presents an \gls{in} for mixed traffic of event and non-event messages. By utilizing a hybrid error and flow control scheme, event messages can be transmitted over the same physical link as reliable non-event messages with low latency / jitter.
For simple traffic patterns, the simulation-based verification validates the design choices and shows that event traffic is transported, absent congestion, with deterministic latency, while the error control scheme for non-event messages recovers quickly from transmission errors and allows throughput close to the link bandwidth with an MTBF of \(\geq\mathcal{O}(\SI{1}{\day})\). In simulation and for a physical transmission delay of \SI{4}{\nano\second} a round trip latency of $\geq$~\SI{32}{\nano\second} is achieved, or \SI{16}{\nano\second} for a single hop transmission and a single router traversal. Note that the router for event messages used here is a simple broadcast router; in a final deployed system a multicast capable event message router is expected to incur an additional overhead.

\cref{tab:otherSystems} gives a comparison with the \gls{in} of other systems. Note, this is a simplified representation in many cases as a simple table cannot capture all details of the respective \gls{in}s. Readers are referred to the cited works to find a more detailed description of the respective \gls{in}.

First we compare to previously proposed \gls{in}s for the \gls{bss2} \gls{asic}. \textcite{thommes2022demonstrating} demonstrate an \gls{in} utilizing the EXTOLL network. It inherits its event and non-event flow and error control from it, which is credit counting and \gls{arq} based. Unlike this work, events are packed into larger packets to minimize the overhead of packets and transmitted with a timestamp to compensate for the jitter introduced by the packing of the events. This introduces additional latency on the receiver side and introduces overhead, as the timestamp has to be transmitted.
The \gls{in} presented in \textcite{ilmberger2025multichip} uses a central \gls{fpga}-based router to connect up to \num{12} \gls{bss2} chips together in a star topology. Event messages are transported asynchronously without timestamps, similar to the approach presented here, but no scheme for transporting non-event messages is described.

Comparing to other neuromorphic systems, these can be split into two groups, systems utilizing global synchronization like Loihi and TrueNorth and systems operating in a globally asynchronous fashion. The \gls{in} implemented by the SpiNNaker-1 \gls{c2c} operates in an asynchronous fashion without further error control or detection. A handshake is used for flow control and on congestion packets are inserted into a retry buffer. Reliable communication between cores or chips is implemented in software for SpiNNaker-1.
Across \gls{pcb} edges SpiNNaker-1 uses spiNNlink, which uses a hybrid credit counting and per channel flow control scheme and an ARQ protocol for error control of both event and non-event messages. Finally the DYNAP-SE2 \gls{in} operates in a similar fashion to the SpiNNaker-1 \gls{c2c}-based \gls{in} utilizing an asynchronous physical link with handshake for flow control.

The primary remaining challenge of the \gls{in} proposed in this work is a routing scheme for the event messages optimized for neuromorphic workloads, especially natively handling multicast messages. 
Finally the designed \gls{in} is optimized for usage with the \gls{bss2} neuromorphic system. Extending the design to longer event messages composed of multiple \gls{phit}s and by supporting a third class of traffic, messages for which errors are detected but not corrected would make the architecture more applicable to a wider range of neuromorphic systems.
 \printbibliography

\end{document}